\documentclass[prl,aps,twocolumn,showpacs,floatfix,superscriptaddress]{revtex4-1}
\usepackage{soul}
\usepackage{graphicx}
\usepackage{dcolumn}
\usepackage{bm}
\usepackage{amsmath}
\usepackage{amssymb}
\usepackage{color}
\usepackage{hyperref}
\hypersetup{
    colorlinks=true,
    linkcolor=blue,
    filecolor=blue,
    urlcolor=blue,
    citecolor=blue,
}

\begin{document}

\title{Emergence of Chiral Dynamical Multiferroicity in a  Ferroelectric Lattice Nonadiabatically Driven by Ultrafast Achiral Electric Fields}

\author{Chuanbao Zhang}
\affiliation{International Center for Quantum Design of Functional Materials (ICQD), Hefei National Research Center for Physical Sciences at the Microscale, University of Science and Technology of China, Hefei 230026, China}
\affiliation{Hefei National Laboratory, University of Science and Technology of China, Hefei 230088, China}

\author{Zhizhong Ding}
\affiliation{International Center for Quantum Design of Functional Materials (ICQD), Hefei National Research Center for Physical Sciences at the Microscale, University of Science and Technology of China, Hefei 230026, China}
\affiliation{Hefei National Laboratory, University of Science and Technology of China, Hefei 230088, China}

\author{Ping Cui}
\thanks{Corresponding author:\\cuipg@ustc.edu.cn}
\affiliation{International Center for Quantum Design of Functional Materials (ICQD), Hefei National Research Center for Physical Sciences at the Microscale, University of Science and
Technology of China, Hefei 230026, China}
\affiliation{Hefei National Laboratory, University of Science and Technology of China, Hefei 230088, China}

\author{Zhenyu Zhang}
\thanks{Corresponding author:\\zhangzy@ustc.edu.cn}
\affiliation{International Center for Quantum Design of Functional Materials (ICQD), Hefei National Research Center for Physical Sciences at the Microscale, University of Science and Technology of China, Hefei 230026, China}
\affiliation{Hefei National Laboratory, University of Science and Technology of China, Hefei 230088, China}

\date{\today}

\begin{abstract}
It has been shown recently that chiral dynamical multiferroicity can be generated on a ferroelectric lattice whose electric dipoles respond masslessly under chiral optical pumping.
Here we demonstrate that, when driven by ultrafast electric pulses, chiral dynamical multiferroicity can also emerge even in situations where the external field is achiral.
We reveal this striking phenomenon using a prototypical system of a BaTiO$_3$ moir\'e ferroelectric lattice, emphasizing the key factor that its chiral electric dipoles inevitably behave massively upon ultrafast driving.
At a deeper level, the massive nature is attributed to the anisotropic and nonadiabatic dipolar responses, as captured by the phase difference between the faster longitudinal and slower transverse components of the ferroelectric polarization.
Crucially, such a phase difference naturally also gives rise to dynamical magnetization, which exhibits chiral magnetic textures with monopole-like topology coexisting with the ferroelectric chirality.
These findings establish the dipolar mass as an enabling and tunable degree of freedom in inducing chiral dynamical multiferroicity, offering new avenues for ultrafast, non-contact magnetic control using pure electric probes.
\end{abstract}

\maketitle
Effective coupling between ferroelectric polarization and magnetization lies at the core of multiferroics, offering a pathway to control magnetism via electric fields or vice versa for a broad range of applications \cite{Eerenstein2006, Cheong2007, Wang2009, Tokura2014, Fiebig2016, Spaldin2019}.
While traditional multiferroicity is constrained to the static regime with stringent symmetry requirements, dynamical multiferroicity has recently been proposed as an appealing new concept, where a time-varying electric polarization $\mathbf{P}$ can induce a net magnetization $\mathbf{M}$ in the form of $\mathbf{M} \propto \mathbf{P} \times \dot{\mathbf{P}}$ \cite{Juraschek2017}. In essence, this particular form defining the finite $\mathbf{M}$ is formally analogous to the existence of an effective ferroelectric Dzyaloshinskii$-$Moriya coupling, which in turn can result in chiral excitations such as ferroelectric skyrmions \cite{Zhao2021DMlike, Chen2022eDMI, Junquera2023RMP}. This rationale suggests that the creation of a dynamical multiferroic state can be achieved by driving the atoms in a lattice into gyrating trajectories with finite angular momenta that inherently lead to ferroelectric chirality. Indeed, theoretical proposals and experimental realizations have predominantly relied on chiral optical pumping (e.g., using circularly polarized pulses) to explicitly imprint the necessary angular momenta onto the lattice excitations \cite{Nova2017,  Geilhufe2021, Juraschek2022, Basini2024, Paiva2025}.
Recent examples have further shown that vortexed optical beams carrying orbital angular momentum can also induce dynamical multiferroicity and magnetic topological structures in intrinsically nonmagnetic ferroelectrics \cite{Gao2023PRL, Gao2024PRL, Gao2024PRB}.

In the realm of dynamical phenomena, a central physical ingredient is the inertia in the responses of a system upon time-dependent external probing.
For magnetic systems, the role of  inertia has recently garnered significant attention, where the effective mass of the spin can result in nutational motion superimposed on the conventional precessional dynamics \cite{Kimel2009, Bhattacharjee2012, Neeraj2021, Unikandanunni2022, Rodriguez2024}.
Such an inertial term is essential in properly describing the spin dynamics at the femtosecond timescale. For ferroelectrics, inertial effects associated with domain-wall motion and soft-mode dynamics have also been shown to play an important role under ultrafast driving \cite{Akamatsu2018, Yang2020, Li2021, Stoica2024, Li2025, Wang2025}. Yet, so far theoretical and experimental demonstrations of dynamical multiferroicity have been mainly limited to the massless regime of ferroelectric dynamics, even though the systems are under ultrafast optical driving.

In this Letter, we demonstrate that chiral dynamical multiferroicity can be generated by applying a strictly achiral ultrafast electric pulse when the inertial effect is inevitably included. Using phase-field simulations of a prototypical twisted BaTiO$_3$ moir\'e ferroelectric lattice, we identify the dipolar mass as the enabling degree of freedom that induces dynamical magnetization.
We show that the system enters a distinct nonadiabatic regime, where the anisotropy in the inertial responses creates a phase difference between the faster longitudinal and slower transverse components of the dipolar motion. Such a nutational phase delay spontaneously generates dynamical magnetization, transforming the static polar vortices into chiral magnetic textures with monopole-like topology. Our results identify dipolar mass as an inherent physical ingredient in inducing chiral dynamical multiferroicity, enabling ultrafast magnetic recording with a linearly polarized electric pulse.

\begin{figure}[t]
\includegraphics[width=1.0\columnwidth]{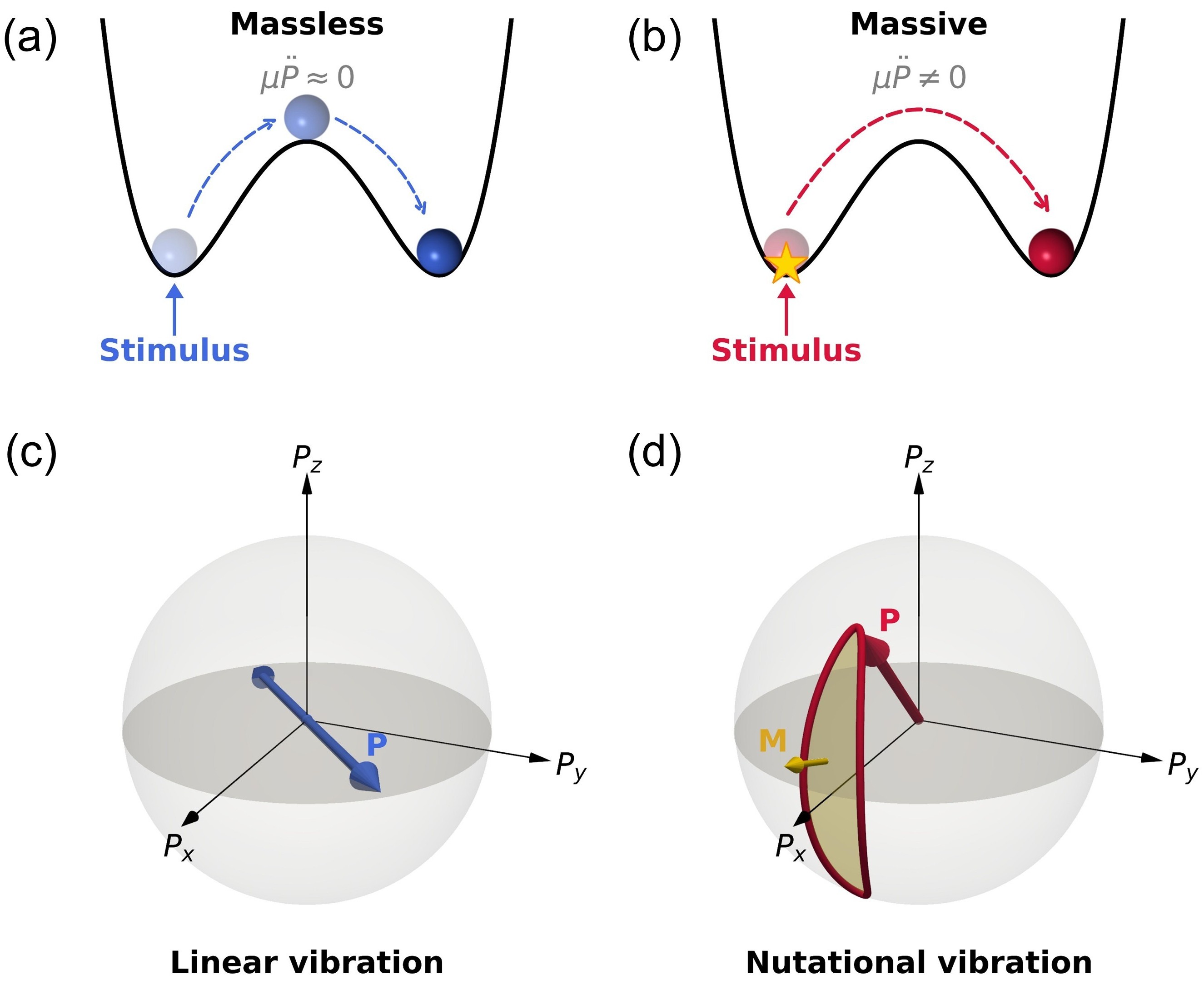}
\caption{\label{fig:mechanism}(a) In the massless regime ($\mu \ddot{\mathbf{P}} \approx 0$), the polarization relaxes along the steepest descent path (blue dashed), resulting in a reversible trajectory with zero net magnetization.
(b) In the massive regime ($\mu \ddot{\mathbf{P}} \neq 0$), the system exhibits ballistic overshoot (red dashed), enabling access to nonequilibrium states.
(c) Without inertial coupling, a linear drive leads to linear vibrations on the order-parameter sphere.
(d) With inertial coupling, the polarization vector undergoes nutation, tracing a loop that encloses a finite solid angle (gold shaded area), thus generating a dynamical magnetization $\mathbf{M}$ (gold arrow) perpendicular to the loop.}
\end{figure}

The distinction between the conventional massless (or overdamped) dynamics and the massive inertial dynamics is conceptually illustrated in Fig.~\ref{fig:mechanism}.
In the overdamped limit [Fig.~\ref{fig:mechanism}(a)], commonly described by the time-dependent Ginzburg-Landau equation $\gamma \dot{\mathbf{P}} = -\delta F/\delta \mathbf{P}$, where $\gamma$ is the damping coefficient and $F$ is the total free energy, the system behaves like a massless particle in a viscous fluid \cite{Wang2019}.
Under a linear electric pulse, the polarization moves and returns along the same trajectory. On the order-parameter sphere, this massless response reduces to a purely linear back-and-forth vibration [Fig.~\ref{fig:mechanism}(c)], yielding $\oint \mathbf{P} \times d\mathbf{P} = 0$.
Conversely, when the kinetic energy term $\frac{1}{2}\mu |\dot{\mathbf{P}}|^2$ is included [Fig.~\ref{fig:mechanism}(b)], the governing equation becomes the inertial Landau-Khalatnikov equation: $\mu \ddot{\mathbf{P}} + \gamma \dot{\mathbf{P}} = -\delta F/\delta \mathbf{P}$, where $\mu$ is the effective dipolar mass parameter.
Here, the momentum of the inertial polarization allows it to overshoot the energy maximum, breaking the trajectory reversibility.
An electric field applied along the $z$-axis ($\mathbf{E} \parallel \hat{z}$) drives $P_z$ directly.
However, the Landau potential contains biquadratic coupling terms, most notably $\alpha_{12} P_z^2 (P_x^2 + P_y^2)$, which acts as a time-dependent stiffness for the transverse components.
In the massless limit, $P_x$ and $P_y$ would instantaneously adjust to the changing $P_z$, maintaining the system at the instantaneous energy minimum.
In the massive regime, however, the transverse response is governed by its own inertial timescale $\tau_{\perp} \approx \sqrt{\mu / (\partial^2 F / \partial P_\perp^2)}$, which differs from the driving longitudinal timescale.
This frequency mismatch creates a retardation effect such that $P_z(t)$ reaches its maximum before $P_{x}(t)$ and $P_{y}(t)$  can fully relax; during the recovery phase, the inertia drives the system along a different return path.
As visualized in Fig.~\ref{fig:mechanism}(d), this results in a nutational trajectory on the order-parameter sphere, conceptually analogous to the nutational motion of a symmetric top.
The open loop in phase space signifies a finite angular momentum $\mathbf{L}$, and consequently, a transient magnetization $\mathbf{M} \propto \mathbf{L}$.
In essence, the nutation is enabled by an inertia-induced phase lag between the coupled polarization components, which makes $\mathbf{P}$ and $\dot{\mathbf{P}}$ noncollinear. Equivalently, the cycle-averaged magnetization is proportional to the oscillatory loop area in the phase plane, as detailed in Section S1 of Supplemental Material (SM) \cite{SM}.

The nutational mechanism outlined above can be viewed as an inertial parametric oscillator, where the linear drive pumps energy into the orthogonal modes via the nonlinear coupling, a key process missing in the overdamped limit \cite{nonlinear2011, Subedi2014}.
Unlike the inverse Faraday effect, where a chiral drive is required to inject an axial angular momentum \cite{vanDerZiel1965, Mironov2021, Majedi2021}, here no external chirality is required. The handedness of the response is selected by the static structural chirality of the moir\'e texture, while the $z$-polarized achiral pulse provides the longitudinal drive. Crucially, as demonstrated below, the dipolar mass induces an inertial phase lag that unlocks transverse motion and renders $\mathbf{P}$ noncollinear with $\dot{\mathbf{P}}$, thereby generating $\mathbf{M}\propto \mathbf{P}\times \dot{\mathbf{P}}$. To quantify this effect in a realistic material, we perform phase-field simulations of a twisted multilayer BaTiO$_3$ system under a driving pulse (see Sections S2 and S3).
The moir\'e interference pattern generates a periodic strain and flexoelectric field, stabilizing a lattice of polar vortices and anti-vortices \cite{SanchezSantolino2024, Lee2024, Prosandeev2025}.
We subject this lattice to an ultrafast Gaussian electric pulse $E_z(t)$ with a duration of 1 ps, polarized along the $z$-axis.

\begin{figure}[t]
\includegraphics[width=1.0\columnwidth]{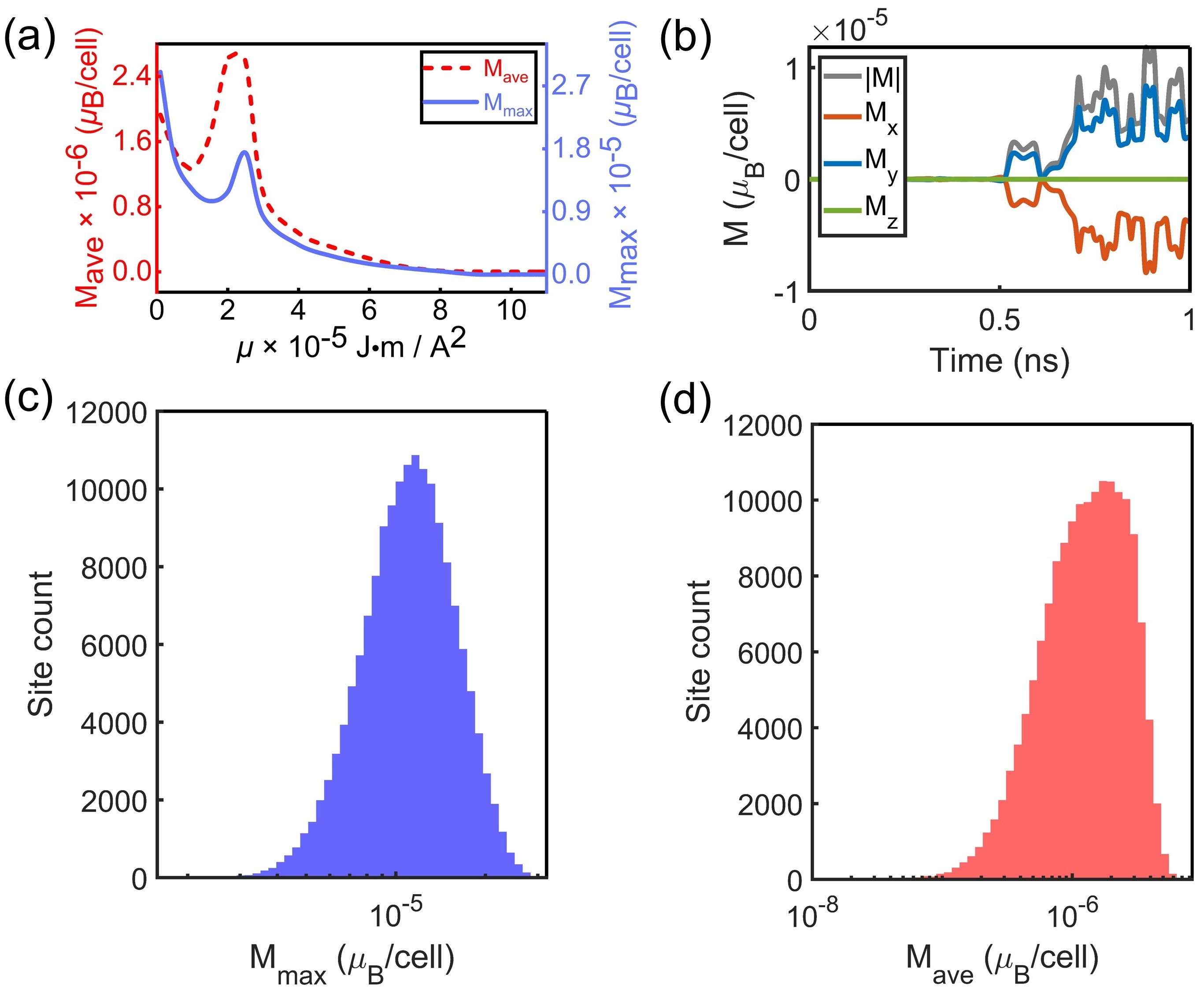}
\caption{\label{fig:time_evolution}(a) Resonance of the maximum ($M_{\rm max}$; blue) and average ($M_{\rm ave}$; red) magnetizations as functions of the inertial parameter $\mu$.
The peaks indicate inertial resonance conditions. (b) Time-resolved magnetization components within a unit cell.
While $M_z$ (green) remains zero due to symmetry, large in-plane components $M_x$ (red) and $M_y$ (blue) emerge dynamically. (c),(d) Histograms of $M_{\rm max}$ and $M_{\rm ave}$ across the entire moir\'e superlattice.}
\end{figure}

The results are shown in Fig.~\ref{fig:time_evolution}(a), where we observe a pronounced resonance peak in the magnetization around
$\mu \approx 2.5 \times 10^{-11}$~J$\cdot$m/A$^2$, as most clearly reflected in the time-averaged magnetization
$M_{\rm ave}$ (and also visible in the maximum value $M_{\max}$).
This resonance occurs when the inertial timescale of the polarization oscillation,
$\tau \sim \sqrt{\mu/|\alpha|}$, becomes comparable to the intrinsic timescale set by the pulse spectrum
and the subsequent lattice relaxation, thereby enabling an underdamped nutational response with a large
phase lag and a finite loop area in the polarization phase space.
Here we note that the sizable magnetization observed as $\mu \to 0$ in the full phase-field simulations, unlike the overdamped-limit case shown in Fig.~\ref{fig:mechanism}(a) (which can be demonstrated numerically using a simplified 0D model; see Fig.~S1), does
not originate from the inertial phase lag; rather, in the moir\'e vortex lattice, a preexisting
transverse polarization texture $P_{\perp}(\mathbf{r}) \neq 0$ converts the ultrafast rise/fall of the
driving component $P_z(t)$ into a short-lived pulse-edge spike
($M \sim -\gamma_{\rm mag}\, P_{\perp}\dot{P}_z$), which can dominate the global maximum $M_{\max}$ in the overdamped limit.
As $\mu$ increases from zero, this edge-following spike is initially suppressed because finite inertia
smooths $\dot{P}_z$, whereas the genuinely inertial contribution
$M \sim \gamma_{\rm mag}\, P_z \dot{P}_{\perp}$ is activated and resonantly enhanced at intermediate
$\mu$, producing the observed peak and the strong nonmonotonic trend.
The sharpness of the resonance further indicates an underdamped inertial mode, suggesting that once excited, the nutational motion can persist for multiple cycles before
damping out.

Figure~\ref{fig:time_evolution}(b) displays the time evolution of the dynamical magnetization components at a representative single-unit-cell site within a polar vortex.
Remarkably, although the drive is achiral ($E_z$), significant in-plane magnetization components $M_x$ and $M_y$ are generated, reaching magnitudes of $\sim 10^{-5} \mu_B$/unit cell.
The $z$-component $M_z$ remains negligible, consistent with the preservation of rotational symmetry in the $xy$-plane by the drive along $z$.
This observation confirms that the magnetization arises from the coupling between the driving $z$-mode and the driven transverse fluctuations.
It is instructive to compare quantitatively with recent experimental and theoretical reports of phonon magnetic moments and phonon-driven magneto-optical signals in nonmagnetic perovskites such as SrTiO$_3$ and KTaO$_3$ \cite{Geilhufe2021, Juraschek2022, Basini2024}.
In SrTiO$_3$, circularly polarized terahertz driving produces a clear helicity-dependent, time-resolved Kerr response, and the purely ionic moment scale is on the order of the nuclear magneton ($\sim 10^{-3}\mu_B$/unit cell) \cite{Basini2024}.
More broadly, time-resolved Faraday rotation and Kerr ellipticity measurements have resolved phonon-driven transient magnetization in other materials platforms \cite{Luo2023,Davies2024}.
While our induced moments are smaller, conceptually, it should be noted that here such chiral dynamical magnetizations are achieved with a linearly polarized pulse without requiring complex chiral optical setups. When the pulse is increased to the THz regime, the resulting magnetization should be substantially enhanced in magnitude as well.
Moreover, even with the magnitude of $10^{-5}\,\mu_B$/unit cell, it should be detectable with state-of-the-art ultrafast Kerr/Faraday polarimetry, especially when combined with balanced detection, long averaging, and/or optical enhancement schemes that are routinely used in ultrafast magneto-optical measurements \cite{Kirilyuk2010}.

The histograms in Figs.~\ref{fig:time_evolution}(c) and~\ref{fig:time_evolution}(d) demonstrate that the emergence of dynamical magnetization is not an isolated local event but a robust collective behavior across the entire moir\'e superlattice.
The broad distribution reflects the heterogeneity of the moir\'e potential, where different spatial regions possess slightly different effective stiffnesses and thus different local resonance conditions.
This inhomogeneous broadening is actually beneficial for experimental observation, as it ensures that a broadband pulse can excite magnetization across a large portion of the sample.
\begin{figure}[t] \includegraphics[width=1.0\columnwidth]{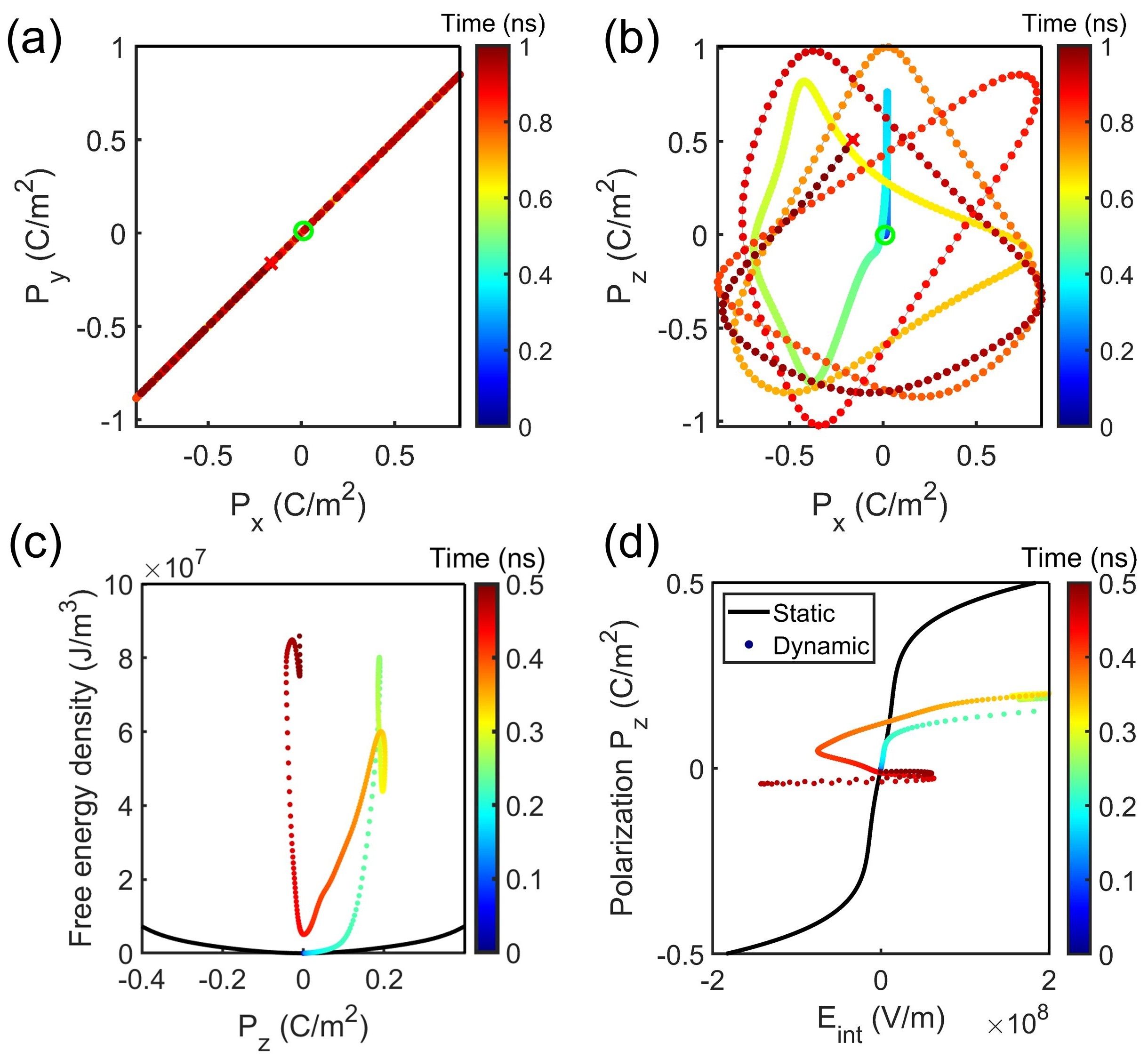} \caption{\label{fig:phase_space}(a) Linear dependence of the in-plane $P_y$-$P_x$ trajectory, indicating their in-phase oscillations and zero $M_z$. The green circle marks the initial time, and the red cross marks the final time. (b) Nonlinear dependence of the $P_z$-$P_x$ trajectory, forming a butterfly loop, and revealing the phase lag caused by inertia that generates $M_y$. (c) Evolution of the free energy density, with the high energy states (red) reached by the ballistic force. (d) Dynamical $P_z$-$E_{int}$ hysteresis loop.} \end{figure}

The geometric origin of the magnetization is visualized in the phase space (Fig.~\ref{fig:phase_space}).
The trajectory in the $P_x$-$P_y$ plane [Fig.~\ref{fig:phase_space}(a)] is essentially linear, enclosing zero area, which explains the absence of $M_z$.
In contrast, the trajectories in the $P_z$-$P_x$ [Fig.~\ref{fig:phase_space}(b)] plane exhibit large, open loops resembling a butterfly shape.
The areas enclosed by these loops are proportional to the generated magnetizations ($\int M_y dt \propto \oint P_z dP_x$). In this sense, the polarization coordinate executes an orbital-angular-momentum$-$like motion, conceptually analogous to chiral phonons, where circular (elliptical) ionic trajectories with a definite handedness generate an orbital magnetic moment \cite{Zhang2015, Zhu2018, Ueda2023, Ishito2023, Juraschek2025}. Here, the handedness is not imposed by an external chiral drive but emerges from the inertial phase delay between the driving longitudinal component and the transverse responses.
These loops arise because the driving mode $P_z$ responds adiabatically to the field, while the transverse components $P_x$ and $P_y$ respond nonadiabatically with an inertial delay. Thermodynamically, such a process involves driving the system away from its equilibrium state.

Figure~\ref{fig:phase_space}(c) shows the evolution of the free energy with increasing the field strength $E_{int}$ of the pulse.
The system surmounts the local energy barriers, accessing states that are otherwise inaccessible in the adiabatic limit. Intriguingly, Fig.~\ref{fig:phase_space}(d) highlights a dynamic negative capacitance (NC) effect \cite{Salahuddin2008, Khan2011, Appleby2014, Khan2015, Iniguez2019, Hoffmann2019}.
In static equilibrium, the polarization slope is positive.
However, under ballistic driving, the inertial term $\mu \ddot{\mathbf{P}}$ provides a kinetic overshoot channel, allowing the system to transiently traverse this inverted-curvature region that is thermodynamically unstable.
Accordingly, the trajectory (colored dots) deviates significantly from the static equilibrium curve (black line).
During this NC interval, energy pumped from the external field is preferentially stored in the kinetic channel associated with the dipolar mass, facilitating the large-amplitude nutational motion and enlarged phase-space loop area that underlie the dynamical magnetization.
Such behavior is reminiscent of dynamical stabilization in Kapitza pendulums \cite{Citro2015}, where rapid driving can stabilize an otherwise unstable configuration.
Here, the inertial dynamics allows the ferroelectric dipole to act as a kinetic capacitor, temporarily storing energy in the nutational motion of the order parameter, while the associated orbital motion generates the dynamical magnetization (as detailed in Section S5).

\begin{figure}[t]
\includegraphics[width=1.0\columnwidth]{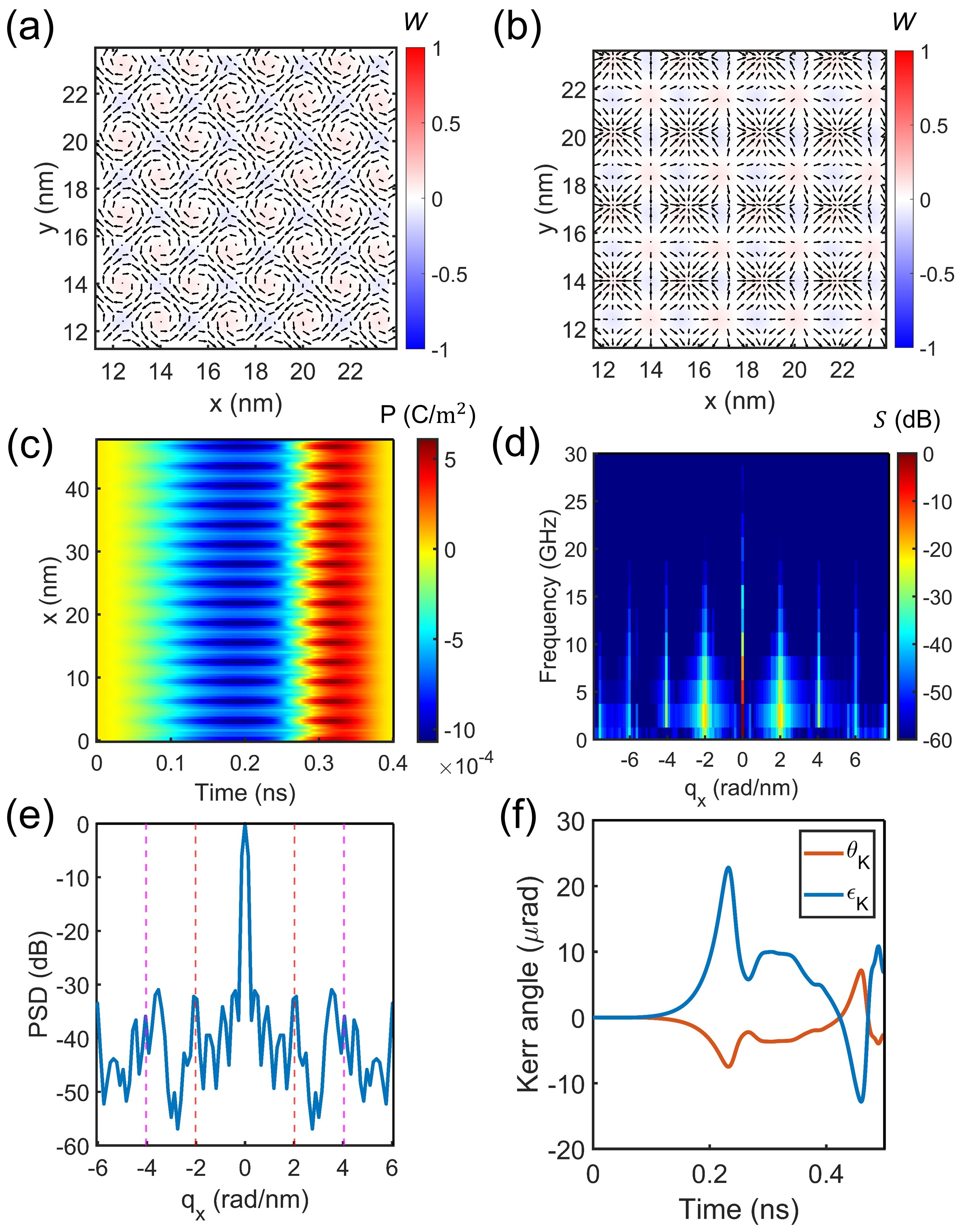}
\caption{\label{fig:moire_topology}(a) A static moir\'e polar vortex lattice showing $\mathbf{P}_{xy}$ vectors and winding numbers $W$ (color).
(b) Dynamical magnetization vectors $\mathbf{M}_{xy}$ and their winding numbers $W$ (color). (c) Space-time diagram of $P_x$ showing coherent collective oscillations of the moir\'e lattice. (d) Dynamical structure factor $S(q_x, \omega)$. (e) Simulated SHG patterns as measured by the power spectral density (PSD); the dashed lines mark the values at $\pm q_{\text{moir\'e}}$ and $\pm 2q_{\text{moir\'e}}$. (f) Time-resolved magneto-optic Kerr effect (TR-MOKE), with the Kerr rotation angle $\theta_K$ and ellipticity $\epsilon_K$ constructed from the simulated dynamical magnetization.}
\end{figure}

Finally, we explore how this inertial mechanism interplays with the chirality of the moir\'e lattice.
The static ground state of the twisted BaTiO$_3$ multilayer consists of a periodic array of polar vortices and anti-vortices [Fig.~\ref{fig:moire_topology}(a)], characterized by a nonzero curl $\nabla \times \mathbf{P}$.
Upon driving with the pulse along $z$, we observe a chiral heritage effect in which the curl-carrying electric vortices are mapped onto the divergence-carrying magnetic textures, yielding $\nabla \cdot \mathbf{M} \neq 0$. This mapping can be understood geometrically through the vector product relation (Section S6). Consequently, the ferroelectric vortices in Fig.~\ref{fig:moire_topology}(a) induce the accompanying magnetic monopoles seen in Fig.~\ref{fig:moire_topology}(b).
Such a chiral imprinting implies that the topological charge of the electric texture determines the magnetic charge of the dynamical state.
Specifically, an electric vortex with winding number $+1$ generates a magnetic source (divergence $>0$), while an anti-vortex with winding number $-1$ generates a magnetic sink (divergence $<0$). Such a phenomenon provides a selection rule for writing magnetic textures via electric topology, offering a novel route to creating dynamical magnetic monopole lattices.

The collective nature of the dipolar response is also evident in the space-time diagram [Fig.~\ref{fig:moire_topology}(c)], which shows coherent oscillations of the polarization field propagating through the lattice.
The stripe patterns indicate that the moir\'e lattice oscillates coherently as a whole, rather than as isolated dipoles.
To characterize the elementary excitations in such a dynamical system, we compute the dynamical structure factor $S(q_x, \omega)$ [Fig.~\ref{fig:moire_topology}(d)].
The spectrum reveals well-defined, discrete dispersive branches. These modes simultaneously carry electric polarization and dynamical magnetization. Crucially, our full three-dimensional simulations reveal that the observed chirality imprinting is not merely a surface effect but exhibits a complex layer-dependent behavior determined by the moir\'e geometry (Fig. S2). The twisted architecture also imposes a structural chirality flip across the sample thickness.
Specifically, the electric vortex chirality in the bottom layers (Layers 1-5) is opposite to that in the top layers (Layers 7-11) (Figs. S3 and S4). The temporal robustness of these layer-dependent dynamics is further confirmed in Fig. S5.

Before closing, we comment on experimental observables that can directly map the coupled polar and magnetic dynamics predicted here.
Figure~\ref{fig:moire_topology}(e) shows a second-harmonic generation (SHG) diffraction signature of the moir\'e polar texture. Specifically, the power spectral density (PSD) of the SHG intensity along the $q_x$ axis  exhibits satellite features near $\pm q_{\text{moir\'e}}$ and higher harmonics near $\pm 2q_{\text{moir\'e}}$, providing a direct nonlinear-optical fingerprint of the moir\'e periodicity and its nonsinusoidal modulation. Complementarily, Fig.~\ref{fig:moire_topology}(f) presents a time-resolved magneto-optic Kerr effect (TR-MOKE) proxy constructed from the simulated dynamical magnetization, in which the Kerr rotation/ellipticity maps the ultrafast buildup, overshoot, and sign reversals of the induced magnetic response. Together, the SHG and TR-MOKE measurements offer a practical, dual-channel roadmap to experimentally verify the inertial chiral dynamical multiferroicity and the electric-to-magnetic chiral texture transfer reported in this work. Recent work has predicted that uniform optical illumination can induce spatially nonuniform charge redistribution within a moiré supercell \cite{Guo2026}. Our finding that an achiral ultrafast electric pulse can generate a chiral dynamical magnetization provides a complementary manifestation of a broader motif, namely that intrinsic nonlinearity in ordered media can convert a symmetry-simple drive into a symmetry-nontrivial response across distinct physical channels. More broadly, once inertia becomes relevant, texture dynamics is no longer purely diffusive but acquires a wave-like character with an effective finite propagation speed; in this limit, one may even expect observations of non-relativistic to relativistic transitions of domain-wall kinematics, as detailed in Section S11.

In conclusion, we have demonstrated that the inherent mass of ferroelectric dipoles serves as an enabling degree of freedom that permits the generation of chiral dynamical multiferroicity using ultrafast achiral electric fields.
In the regime of nonnegligible inertia, the anisotropic response of the lattice leads to nutational trajectories that in turn break time-reversal symmetry and yield dynamical magnetization.
In moir\'e ferroelectric superlattices, such a process facilitates the ballistic imprinting of chirality into the magnetic textures.
Our work suggests that inertial dynamics can be harnessed to bypass symmetry constraints in light-matter interactions, paving the way for ultrafast magnetoelectric devices.
\begin{acknowledgments}
This work was supported by the National Natural Science
Foundation of China (Grant Nos. 12374458, 12488101,
and 11974323), the Innovation Program for Quantum Science
and Technology (Grant No. 2021ZD0302800), the Anhui
Initiative in Quantum Information Technologies (Grant
No. AHY170000), the Strategic Priority Research Program of
Chinese Academy of Sciences (Grant No. XDB0510200), and
the Anhui Provincial Key Research and Development Project
(Grant No. 2023z04020008).
\end{acknowledgments}


\begin{thebibliography}{58}
\bibitem{Eerenstein2006} W. Eerenstein, N. D. Mathur, and J. F. Scott, Multiferroic and magnetoelectric materials, \href{https://doi.org/10.1038/nature05023}{Nature \textbf{442}, 759 (2006)}.
\bibitem{Cheong2007} S.-W.\ Cheong and M. Mostovoy, Multiferroics: a magnetic twist for ferroelectricity, \href{https://doi.org/10.1038/nmat1804}{Nat. Mater. \textbf{6}, 13 (2007)}.
\bibitem{Wang2009} K. F. Wang, J.-M. Liu, and Z. F. Ren, Multiferroicity: the coupling between magnetic and polarization orders, \href{https://doi.org/10.1080/00018730902920554}{Adv. Phys. \textbf{58}, 321 (2009)}.
\bibitem{Tokura2014} Y. Tokura, S. Seki, and N. Nagaosa, Multiferroics of spin origin, \href{https://doi.org/10.1088/0034-4885/77/7/076501}{Rep. Prog. Phys. \textbf{77}, 076501 (2014)}.
\bibitem{Fiebig2016} M. Fiebig, T. Lottermoser, D. Meier, and M. Trassin, Evolution of multiferroics, \href{https://doi.org/10.1038/natrevmats.2016.46}{Nat. Rev. Mater. \textbf{1}, 16046 (2016)}.
\bibitem{Spaldin2019} N. A. Spaldin and R. Ramesh, Advances in magnetoelectric multiferroics, \href{https://doi.org/10.1038/s41563-018-0275-2}{Nat. Mater. \textbf{18}, 203 (2019)}.
\bibitem{Juraschek2017} D. M. Juraschek, M. Fechner, A. V. Balatsky, and N. A. Spaldin, Dynamical multiferroicity, \href{https://doi.org/10.1103/PhysRevMaterials.1.014401}{Phys. Rev. Mater. \textbf{1}, 014401 (2017)}.
\bibitem{Zhao2021DMlike} H. J. Zhao, P. Chen, S. Prosandeev, S. Artyukhin, and L. Bellaiche, Dzyaloshinskii--Moriya-like interaction in ferroelectrics and antiferroelectrics, \href{https://doi.org/10.1038/s41563-020-00821-3}{Nat. Mater. \textbf{20}, 341 (2021)}.
\bibitem{Chen2022eDMI} P. Chen, H. J. Zhao, S. Prosandeev, S. Artyukhin, and L. Bellaiche, Microscopic origin of the electric Dzyaloshinskii-Moriya interaction, \href{https://doi.org/10.1103/PhysRevB.106.224101}{Phys. Rev. B \textbf{106}, 224101 (2022)}.
\bibitem{Junquera2023RMP} J. Junquera, Y. Nahas, S. Prokhorenko, L. Bellaiche, J. \'{I}\~{n}iguez, D. G. Schlom, L.-Q. Chen, S. Salahuddin, D. A. Muller, L. W. Martin, and R. Ramesh, Topological phases in polar oxide nanostructures, \href{https://doi.org/10.1103/RevModPhys.95.025001}{Rev. Mod. Phys. \textbf{95}, 025001 (2023)}.
\bibitem{Nova2017} T. F. Nova, A. Cartella, A. Cantaluppi, M. F\"orst, D. Bossini, R. V. Mikhaylovskiy, A. V. Kimel, R. Merlin, and A. Cavalleri, An effective magnetic field from optically driven phonons, \href{https://doi.org/10.1038/nphys3925}{Nat. Phys. \textbf{13}, 132 (2017)}.
\bibitem{Geilhufe2021} R. M. Geilhufe, V. Juri\v{c}i\'c, S. Bonetti, J.-X. Zhu, and A. V. Balatsky, Dynamical multiferroicity in KTaO$_3$, \href{https://doi.org/10.1103/PhysRevResearch.3.L022011}{Phys. Rev. Research \textbf{3}, L022011 (2021)}.
\bibitem{Juraschek2022}
D. M. Juraschek, T. Neuman, and P. Narang,
Giant effective magnetic fields from optically driven chiral phonons in $4f$ paramagnets,
\href{https://doi.org/10.1103/PhysRevResearch.4.013129}{Phys. Rev. Res. \textbf{4}, 013129 (2022)}.
\bibitem{Basini2024} M. Basini \textit{et al.}, Terahertz electric-field-driven dynamical multiferroicity in SrTiO$_3$, \href{https://doi.org/10.1038/s41586-024-07175-9}{Nature \textbf{628}, 534 (2024)}.
\bibitem{Paiva2025} C. Paiva, M. Fechner, and D. M. Juraschek, Dynamically Induced Multiferroic Polarization, \href{https://doi.org/10.1103/7lm1-wm3y}{Phys. Rev. Lett. \textbf{135}, 066702 (2025)}.
\bibitem{Gao2023PRL} L. Gao, S. Prokhorenko, Y. Nahas, and L. Bellaiche, Dynamical multiferroicity and magnetic topological structures induced by the orbital angular momentum of light in a nonmagnetic material, \href{https://doi.org/10.1103/PhysRevLett.131.196801}{Phys. Rev. Lett. \textbf{131}, 196801 (2023)}.
\bibitem{Gao2024PRL} L. Gao, S. Prokhorenko, Y. Nahas, and L. Bellaiche, Dynamical control of topology in polar skyrmions via twisted light, \href{https://doi.org/10.1103/PhysRevLett.132.026902}{Phys. Rev. Lett. \textbf{132}, 026902 (2024)}.
\bibitem{Gao2024PRB} L. Gao, S. Prokhorenko, Y. Nahas, and L. Bellaiche, Effective gyration of polar vortex arrays controlled by high orbital angular momentum of light, \href{https://doi.org/10.1103/PhysRevB.109.L121110}{Phys. Rev. B \textbf{109}, L121110 (2024)}.

\bibitem{Kimel2009} A. V. Kimel, B. A. Ivanov, R. V. Pisarev, P. A. Usachev, A. Kirilyuk, and Th. Rasing, Inertia-driven spin switching in antiferromagnets, \href{https://doi.org/10.1038/nphys1369}{Nat. Phys. \textbf{5}, 727 (2009)}.

\bibitem{Bhattacharjee2012} S. Bhattacharjee, L. Nordstr{\"o}m, and J. Fransson, Atomistic spin dynamic method with both damping and moment of inertia effects included from first principles, \href{https://doi.org/10.1103/PhysRevLett.108.057204}{Phys. Rev. Lett. \textbf{108}, 057204 (2012)}.

\bibitem{Neeraj2021} K. Neeraj, N. Awari, S. Kovalev, D. Polley, N. Zhou Hagstr{\"o}m, S. S. P. K. Arekapudi, A. Semisalova, K. Lenz, B. Green, J.-C. Deinert, I. Ilyakov, M. Chen, M. Bawatna, V. Scalera, M. d'Aquino, C. Serpico, O. Hellwig, J.-E. Wegrowe, M. Gensch, and S. Bonetti, Inertial spin dynamics in ferromagnets, \href{https://doi.org/10.1038/s41567-020-01040-y}{Nat. Phys. \textbf{17}, 245 (2021)}.

\bibitem{Unikandanunni2022} V. Unikandanunni, R. Medapalli, M. Asa, E. Albisetti, D. Petti, R. Bertacco, E. E. Fullerton, and S. Bonetti, Inertial Spin Dynamics in Epitaxial Cobalt Films, \href{https://doi.org/10.1103/PhysRevLett.129.237201}{Phys. Rev. Lett. \textbf{129}, 237201 (2022)}.

\bibitem{Rodriguez2024} R. Rodriguez, M. Cherkasskii, R. Jiang, R. Mondal, A. Etesamirad, A. Tossounian, B. A. Ivanov, and I. Barsukov, Spin Inertia and Auto-Oscillations in Ferromagnets, \href{https://doi.org/10.1103/PhysRevLett.132.246701}{Phys. Rev. Lett. \textbf{132}, 246701 (2024)}.

\bibitem{Akamatsu2018}
H. Akamatsu, Y. Yuan, V. A. Stoica, G. Stone, T. Yang, Z. Hong, S. Lei, Y. Zhu, R. C. Haislmaier, J. W. Freeland, L.-Q. Chen, H. Wen, and V. Gopalan,
Light-Activated Gigahertz Ferroelectric Domain Dynamics,
\href{https://doi.org/10.1103/PhysRevLett.120.096101}{Phys. Rev. Lett. \textbf{120}, 096101 (2018)}.
\bibitem{Yang2020}
T. Yang, B. Wang, J.-M. Hu, and L.-Q. Chen,
Domain Dynamics under Ultrafast Electric-Field Pulses,
\href{https://doi.org/10.1103/PhysRevLett.124.107601}{Phys. Rev. Lett. \textbf{124}, 107601 (2020)}.
\bibitem{Li2021}
Q. Li, V. A. Stoica, M. Pa\'{s}ciak, Y. Zhu, Y. Yuan, T. Yang, M. R. McCarter, S. Das, A. K. Yadav, S. Park, C. Dai, H. J. Lee, Y. Ahn, S. D. Marks, S. Yu, C. Kadlec, T. Sato, M. C. Hoffmann, M. Chollet, M. E. Kozina, S. Nelson, D. Zhu, D. A. Walko, A. M. Lindenberg, P. G. Evans, L.-Q. Chen, R. Ramesh, L. W. Martin, V. Gopalan, J. W. Freeland, J. Hlinka, and H. Wen,
Subterahertz collective dynamics of polar vortices,
\href{https://doi.org/10.1038/s41586-021-03342-4}{Nature \textbf{592}, 376 (2021)}.
\bibitem{Stoica2024}
V. A. Stoica, T. Yang, S. Das, Y. Cao, H. Wang, Y. Kubota, C. Dai, H. Padmanabhan, Y. Sato, A. Mangu, Q. L. Nguyen, Z. Zhang, D. Talreja, M. E. Zajac, D. A. Walko, A. D. DiChiara, S. Owada, K. Miyanishi, K. Tamasaku, T. Sato, J. M. Glownia, V. Esposito, S. Nelson, M. C. Hoffmann, R. D. Schaller, A. M. Lindenberg, L. W. Martin, R. Ramesh, I. Matsuda, D. Zhu, L.-Q. Chen, H. Wen, V. Gopalan, and J. W. Freeland,
Non-equilibrium pathways to emergent polar supertextures,
\href{https://doi.org/10.1038/s41563-024-01981-2}{Nat. Mater. \textbf{23}, 1394 (2024)}.
\bibitem{Li2025}
W. Li, S. Wang, P. Peng, H. Han, X. Wang, J. Ma, J. Luo, J.-M. Liu, J.-F. Li, C.-W. Nan, and Q. Li,
Terahertz excitation of collective dynamics of polar skyrmions over a broad temperature range,
\href{https://doi.org/10.1038/s41567-025-03056-8}{Nat. Phys. \textbf{21}, 1965 (2025)}.
\bibitem{Wang2025}
H. Wang, V. Stoica, C. Dai, M. Pa\'{s}ciak, S. Das, T. Yang, M. A. P. Gon\c{c}alves, J. Kulda, M. R. McCarter, A. Mangu, Y. Cao, H. Padma, U. Saha, D. Zhu, T. Sato, S. Song, M. C. Hoffmann, P. Kramer, S. Nelson, Y. Sun, Q. Nguyen, Z. Zhang, R. Ramesh, L. W. Martin, A. M. Lindenberg, L.-Q. Chen, J. W. Freeland, J. Hlinka, V. Gopalan, and H. Wen,
Terahertz-field activation of polar skyrmions,
\href{https://doi.org/10.1038/s41467-025-64033-6}{Nat. Commun. \textbf{16}, 8994 (2025)}.
\bibitem{Wang2019} J. J. Wang, B. Wang, and L. Q. Chen, Understanding, Predicting, and Designing Ferroelectric Domain Structures and Switching Guided by the Phase-Field Method, \href{https://doi.org/10.1146/annurev-matsci-070218-121843}{Annu. Rev. Mater. Res. \textbf{49}, 127 (2019)}.
\bibitem{SM} See Supplemental Material for details on the phase-field model, material parameters, and additional figures, which includes Refs. [1,25,30,32-37,54-55,59].
\bibitem{Chen2002}
L.-Q. Chen, Phase-field models for microstructure evolution,
\href{https://doi.org/10.1146/annurev.matsci.32.112001.132041}{Annu. Rev. Mater. Res. \textbf{32}, 113 (2002)}.
\bibitem{Wang2004}
J. Wang, S. Q. Shi, L.-Q. Chen, Y. Li, and T.-Y. Zhang, Phase-field simulations of ferroelectric/ferroelastic polarization switching,
\href{https://doi.org/10.1016/j.actamat.2003.10.011}{Acta Mater. \textbf{52}, 749 (2004)}.
\bibitem{Sha2024}
H. Sha, Y. Zhang, Y. Ma, W. Li, W. Yang, J. Cui, Q. Li, H. Huang, and R. Yu,
Polar vortex hidden in twisted bilayers of paraelectric SrTiO$_3$,
\href{https://doi.org/10.1038/s41467-024-55328-1}{Nat. Commun. \textbf{15}, 10915 (2024)}.
\bibitem{Zhang2025}
Y. Zhang, H. Sha, X. Wang, D. Liang, J. Wang, Q. Li, R. Yu, and H. Huang,
Strain-induced moir\'{e} polar vortex in twisted paraelectric freestanding bilayers,
\href{https://doi.org/10.1038/s41535-025-00796-x}{npj Quantum Mater. \textbf{10}, 75 (2025)}.
\bibitem{Li2001}
Y. L. Li, S. Y. Hu, Z. K. Liu, and L.-Q. Chen, Phase-field model of domain structures in ferroelectric thin films,
\href{https://doi.org/10.1063/1.1377855}{Appl. Phys. Lett. \textbf{78}, 3878 (2001)}.
\bibitem{Nettleton1967} R. E. Nettleton, Effective mass of 180$^\circ$ domain wall in single crystal barium titanate, \href{https://doi.org/10.1143/JPSJ.22.1375}{J. Phys. Soc. Jpn. \textbf{22}, 1375 (1967)}.
\bibitem{nonlinear2011} M. F\"orst \textit{et al.}, Nonlinear phononics as an ultrafast route to lattice control, \href{https://doi.org/10.1038/nphys2055}{Nat. Phys. \textbf{7}, 854 (2011)}.
\bibitem{Subedi2014} A. Subedi, A. Cavalleri, and A. Georges, Theory of nonlinear phononics for coherent light control of solids, \href{https://doi.org/10.1103/PhysRevB.89.220301}{Phys. Rev. B \textbf{89}, 220301 (2014)}.
\bibitem{vanDerZiel1965} J. P. van der Ziel, P. S. Pershan, and L. D. Malmstrom, Optically-induced magnetization resulting from the inverse Faraday effect, \href{https://doi.org/10.1103/PhysRevLett.15.190}{Phys. Rev. Lett. \textbf{15}, 190 (1965)}.
\bibitem{Mironov2021} S. V. Mironov, A. S. Mel'nikov, I. D. Tokman, V. Vadimov, B. Lounis, and A. I. Buzdin, Inverse Faraday Effect for Superconducting Condensates, \href{https://doi.org/10.1103/PhysRevLett.126.137002}{Phys. Rev. Lett. \textbf{126}, 137002 (2021)}.
\bibitem{Majedi2021} A. H. Majedi, Microwave-Induced Inverse Faraday Effect in Superconductors, \href{https://doi.org/10.1103/PhysRevLett.127.087001}{Phys. Rev. Lett. \textbf{127}, 087001 (2021)}.
\bibitem{SanchezSantolino2024}
G. S\'{a}nchez-Santolino \textit{et al.},
A 2D ferroelectric vortex pattern in twisted BaTiO$_3$ freestanding layers,
\href{https://doi.org/10.1038/s41586-023-06978-6}{Nature \textbf{626}, 529 (2024)}.

\bibitem{Lee2024}
S. Lee, D. J. P. {de Sousa}, B. Jalan, and T. Low,
Moir\'{e} polar vortex, flat bands, and Lieb lattice in twisted bilayer BaTiO$_3$,
\href{https://doi.org/10.1126/sciadv.adq0293}{Sci. Adv. \textbf{10}, eadq0293 (2024)}.

\bibitem{Prosandeev2025}
S. Prosandeev, C. Paillard, and L. Bellaiche,
Understanding and controlling dipolar Moir\'{e} pattern in ferroelectric perovskite oxide nanolayers,
\href{https://doi.org/10.1103/PhysRevB.111.L180103}{Phys. Rev. B \textbf{111}, L180103 (2025)}.
\bibitem{Luo2023}
J. Luo, T. Lin, J. Zhang, X. Chen, E. R. Blackert, R. Xu, B. I. Yakobson, and H. Zhu,
Large effective magnetic fields from chiral phonons in rare-earth halides,
\href{https://doi.org/10.1126/science.adi9601}{Science \textbf{382}, 698 (2023)}.

\bibitem{Davies2024}
C. S. Davies \textit{et al.},
Phononic switching of magnetization by the ultrafast Barnett effect,
\href{https://doi.org/10.1038/s41586-024-07200-x}{Nature \textbf{628}, 540 (2024)}.

\bibitem{Kirilyuk2010}
A. Kirilyuk, A. V. Kimel, and Th. Rasing,
Ultrafast optical manipulation of magnetic order,
\href{https://doi.org/10.1103/RevModPhys.82.2731}{Rev. Mod. Phys. \textbf{82}, 2731 (2010)}.

\bibitem{Zhang2015}
L. Zhang and Q. Niu, Chiral phonons at high-symmetry points in monolayer hexagonal lattices,
\href{https://doi.org/10.1103/PhysRevLett.115.115502}{Phys. Rev. Lett. \textbf{115}, 115502 (2015)}.

\bibitem{Zhu2018}
H. Zhu \textit{et al.}, Observation of chiral phonons,
\href{https://doi.org/10.1126/science.aar2711}{Science \textbf{359}, 579 (2018)}.

\bibitem{Ueda2023}
H. Ueda \textit{et al.}, Chiral phonons in quartz probed by X-rays,
\href{https://doi.org/10.1038/s41586-023-06016-5}{Nature \textbf{618}, 946 (2023)}.

\bibitem{Ishito2023}
K. Ishito \textit{et al.}, Truly chiral phonons in $\alpha$-HgS,
\href{https://doi.org/10.1038/s41567-022-01790-x}{Nat. Phys. \textbf{19}, 35 (2023)}.

\bibitem{Juraschek2025}
D. M. Juraschek \textit{et al.}, Chiral phonons,
\href{https://doi.org/10.1038/s41567-025-03001-9}{Nat. Phys. \textbf{21}, 1532 (2025)}.

\bibitem{Salahuddin2008} S. Salahuddin and S. Datta, Use of negative capacitance to provide voltage amplification for low power nanoscale devices, \href{https://doi.org/10.1021/nl071804g}{Nano Lett. \textbf{8}, 405 (2008)}.
\bibitem{Khan2011} A. I. Khan, D. Bhowmik, P. Yu, S. Joo Kim, X. Pan, R. Ramesh, and S. Salahuddin, Experimental evidence of ferroelectric negative capacitance in nanoscale heterostructures, \href{https://doi.org/10.1063/1.3634072}{Appl. Phys. Lett. \textbf{99}, 113501 (2011)}.
\bibitem{Appleby2014} D. J. Appleby \textit{et al.}, Experimental observation of negative capacitance in ferroelectrics at room temperature, \href{https://doi.org/10.1021/nl5017255}{Nano Lett. \textbf{14}, 3864 (2014)}.
\bibitem{Khan2015}
A. I. Khan, K. Chatterjee, B. Wang, S. Drapcho, L. You, C. Serrao, S. R. Bakaul, R. Ramesh, and S. Salahuddin,
Negative capacitance in a ferroelectric capacitor,
\href{https://doi.org/10.1038/nmat4148}{Nat. Mater. \textbf{14}, 182 (2015)}.
\bibitem{Iniguez2019}
J. \'{I}\~{n}iguez, P. Zubko, I. Luk'yanchuk, and A. Cano,
Ferroelectric negative capacitance,
\href{https://doi.org/10.1038/s41578-019-0089-0}{Nat. Rev. Mater. \textbf{4}, 243 (2019)}.
\bibitem{Hoffmann2019} M. Hoffmann \textit{et al.}, Unveiling the double-well energy landscape in a ferroelectric layer, \href{https://doi.org/10.1038/s41586-018-0854-z}{Nature \textbf{565}, 464 (2019)}.
\bibitem{Citro2015} R. Citro \textit{et al.}, Dynamical stability of a many-body Kapitza pendulum, \href{https://doi.org/10.1016/j.aop.2015.03.027}{Ann. Phys. \textbf{360}, 694 (2015)}.
\bibitem{Guo2026} R. Guo, H. Chen, W. Duan, Y. Xu, and C. Wang, Manipulating Charge Distribution in Moiré Superlattices by Light, \href{https://doi.org/10.1103/h2k9-7v61}{Phys. Rev. Lett. \textbf{136}, 086001 (2026)}.

\end{thebibliography}

\end{document}